%% file: main.tex
\documentclass[
reprint,
superscriptaddress,
preprintnumbers,
amsmath,
amssymb,
aps,
prl,
floatfix,
]{revtex4-2}

\usepackage{graphicx}
\usepackage{dcolumn}
\usepackage{bm}
\usepackage{subfigure}

\newcommand{\eq}[1]{\begin{equation}#1\end{equation}}
\newcommand{\eqa}[1]{\begin{eqnarray}#1\end{eqnarray}}

\begin{document}

\title{Phase transitions in optimally robust network structures}
\author{Laura Barth}
\email{laurabarthm@gmail.com}
\affiliation{
Helmholtz Institute for Functional Marine Biodiversity (HIFMB), Ammerl\"ander Heerstr. 231, 26129 Oldenburg, Germany}
\affiliation{
Alfred-Wegener Institute (AWI), Helmholtz Center for Polar and Marine Research, Am Handelshafen 12, 27570 Bremerhaven, Germany}
\affiliation{
Carl-von-Ossietzky Universit\"at, Institute for Chemistry and Biology of the Marine Environment (ICBM), Carl-von-Ossietzky Str. 9-11, 26129 Oldenburg, Germany}

\author{Thilo Gross}
\affiliation{
Helmholtz Institute for Functional Marine Biodiversity (HIFMB), Ammerl\"ander Heerstr. 231, 26129 Oldenburg, Germany}
\affiliation{
Alfred-Wegener Institute (AWI), Helmholtz Center for Polar and Marine Research, Am Handelshafen 12, 27570 Bremerhaven, Germany}
\affiliation{
Carl-von-Ossietzky Universit\"at, Institute for Chemistry and Biology of the Marine Environment (ICBM), Carl-von-Ossietzky Str. 9-11, 26129 Oldenburg, Germany}

\date{\today}

\begin{abstract}
If we add links to a network at random, a critical threshold can be crossed where a giant connected component forms. Conversely, if links or nodes are removed at random, the giant component shrinks and eventually breaks. In this paper, we explore which can optimally withstand random removal of a known proportion of nodes. When optimizing the size of the giant component after the attack, the network undergoes an infinite sequence of continuous phase transitions between different optimal structures as the removed proportion of nodes is increased. When optimizing the proportion of links in the giant component, a similar infinite sequence is observed, but the transitions are now discontinuous. 
\end{abstract}

\maketitle

\let\clearpage\relax
\include{sections/1_In}

\include{sections/2_GF}

\include{sections/3_GT}

\include{sections/4_Co}

\bibliography{lit}

\end{document}

%% file: sections/1_In.tex
Complex networks are fascinating due to their broad applicability as a framework to conceptualize real-world systems. Network models can then be used to explore the different phases of the system at hand and identify phase transitions. A prominent example of such a transition is the giant component transition \cite{ER, solomonoff1951connectivity}.
When links are added randomly to a set of nodes, the giant component forms when a critical density of links is reached. Conversely, removing links or nodes at random from a network will eventually break the giant component.

The creation and destruction of giant components were studied extensively in the context of network robustness \cite{albert2000error, callaway2000network, doyle2005robust}. Research in this area has led to an elegant formalism for the computation of giant component sizes in different types of random networks and under various attack scenarios \cite{molloy1998size, newman2001random, berchenko2009emergence}, revisited in \cite{robustness}.

The giant component is large in the sense that it scales with the network size; however, it may nevertheless comprise only a small proportion $s$ of all nodes, depending on the overall number and arrangement of links. The connectivity of the network can be described by the average number of links connecting to a node, the mean degree $z$. If links or nodes are added randomly, the formation of the giant component occurs in a continuous transition, after which $s$ grows continuously with $z$.

The analysis of giant components can be extended to configuration model networks \cite{molloy1995critical}, where links are distributed randomly with the additional constraint that the degree distribution $p_k$ follows a prescribed shape. This distribution specifies the probability that a randomly drawn node has degree $k$.

When links are added using specific rules that involve a lesser degree of randomness, $s$ can change violently \cite{DSouza}. This transition, called \emph{explosive percolation}. Although it was eventually shown that explosive percolation is continuous \cite{dorogovtsev,warnke} it is so steep that can be considered de-facto discontinuous, and its discovery contributed to the identification of other discontinuous transitions \cite{explsync,kurths}.

Here we consider a different question: Suppose we can specify the degree distribution of a configuration model, but we already know that some proportion $r$ of the nodes will be removed in the future. Hence we seek to chose the degree distribution such that the giant component $s$ is as large as possible after this anticipated attack. For simplicity we consider the case where each node has at least one link before the attack, and assume that the total number of links and hence the mean degree $z$ is fixed.   

Below we develop an analytical method to find the optimal degree distribution and show that the optimal distribution is dependent on $r$. When $r$ is small the optimal network is a regular graph, whereas for large $r$ it is a star-like graph. As we change $r$ between these extremes the network undergoes an infinite cascade of continuous phase transitions. A similar cascade is also observed in a related scenario where we maximize the proportion of surviving links that are in the giant component. However in this case the transitions are discontinuous.

%% file: sections/2_GF.tex
We consider a configuration model network defined by the degree distribution $p_k$. For efficient mathematical manipulation, the individual properties can be stored in a generating function $G(x) = \sum_{k=0}^{\infty} p_{k}x^{k}$, where $x$ is a variable that is introduced for the sole purpose of turning the list of numbers into a polynomial, and does not have a direct physical meaning \cite{newman2001random,generatingfunctionology}.
From the generating function, we can compute the mean degree $G'(1) =  z$ and $Q(x) = G'(x)/z$, which generates the mean excess degree distribution $q_k$, i.e.~the probability distribution of arriving at a node with $k$ additional links, when following a random link in a random direction \cite{newman2003structure}. The expectation value of this distribution is the mean excess degree $q = Q'(1)$, the average number of links that we expect to find at the end of a random link.

\begin{figure}[tb]
\includegraphics[width=0.8\columnwidth]{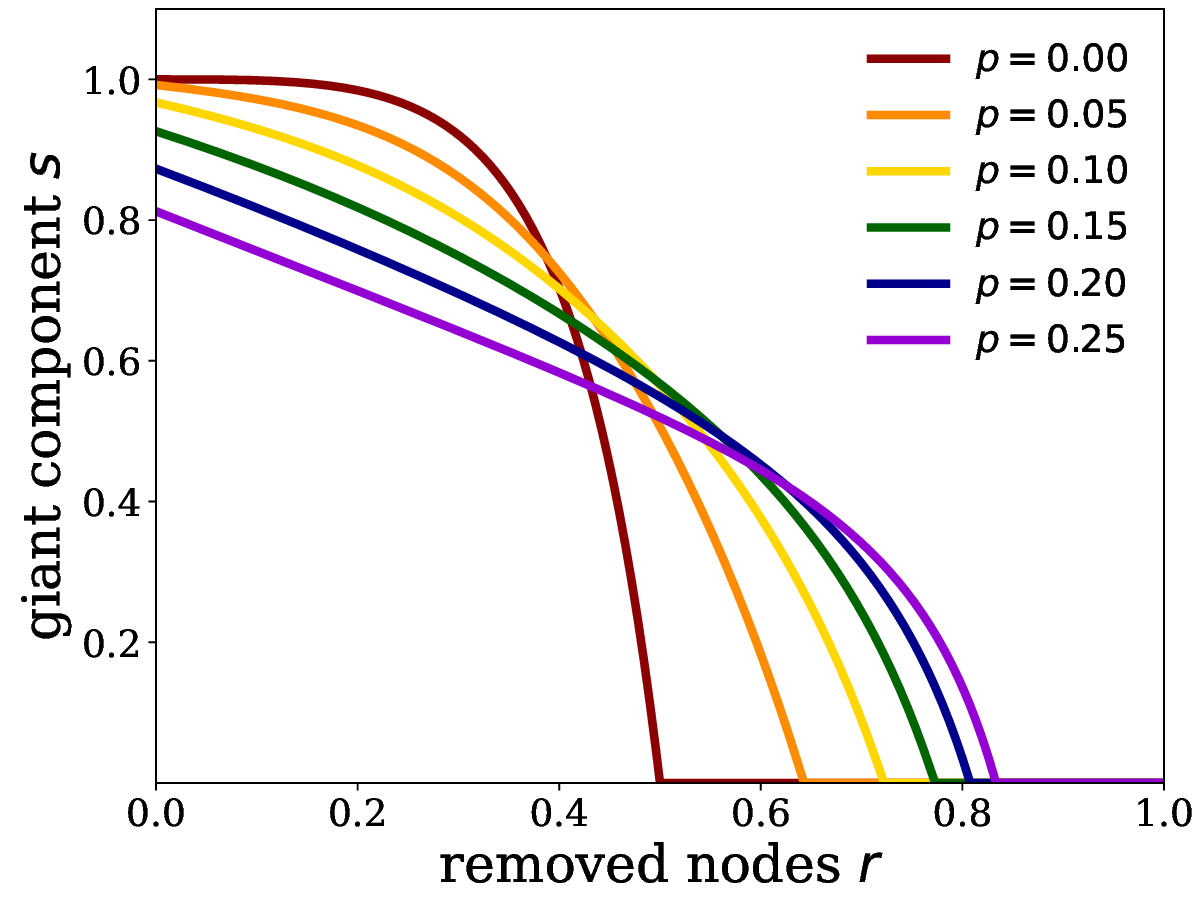}
\caption{\label{sxr}Giant component size after random node removal.
Before the attack, the network contains only nodes of degrees $k = 1, 3, 9$. The heterogeneity of the network is controlled by a single parameter $p$ ranging form $0$ (regular graph) to $0.25$ (maximally heterogeneous). Lines show the size of the giant component $s$ after the removal of a proportion $r$ of the nodes.  This illustrates that the optimal degree distribution depends on anticipated proportion of node removal.
}
\end{figure}

We then consider an `attack' on the network where a proportion $r$ of nodes are removed at random, leaving a proportion $c=1-r$ of surviving nodes. After the attack, the surviving network is described by 
$G_{\rm a}(x) = G(A)$, where $A(x) = cx + r$ \cite{robustness}. 
Hence after the attack, we have a mean degree $z_{\rm a} = G_{\rm a}'(1) = zc$, an excess degree generating function $Q_{\rm a} = Q(A)$, and a mean excess degree $q_{\rm a} = Q_{\rm a}'(1) = qc$.
In configuration model networks, the giant component breaks if $q_{\rm a}<1$ \cite{molloy1995critical, newman2001random}. Hence the corresponding critical point is $r = (q - 1)/q$. Because $q$ is related to the variance of the degree distribution $\sigma^2$ as $q=z+\sigma^2/z$, maximally heterogeneous structures maximize $q$ and hence allow the giant component to withstand the highest proportion of random node removal.  

Here we seek to maximize the size $s$ of the giant component after an attack of known size $r$. To compute $s$, it is useful to introduce $v$ as the probability that following a random link does \emph{not} lead to a node connected to the giant component through other links.
Following \cite{newman2001random,robustness}, the size of the giant component can then be computed by solving 
\eqa{
\label{v}
v &=& Q_{\rm a}(v), \\
\label{s}
s &=& 1 - G_{\rm a}(v).
}
To gain some intuition, it's instructive to first consider networks with $z=3$, where we only permit nodes of the degrees 1, 3 and 9, such that the degree distribution contains only three variables, $p_1$, $p_3$ and $p_9$. Two of these variables are fixed by the normalization condition, $p_1+p_3+p_9=1$, and the constraint on the mean degree, $p_1+3p_3+9p_9=3$. These constraints are met by setting $p_1=3p$, $p_3=1-4p$, $p_9=p$, where $p$ is the remaining degree of freedom, and $p=0$ corresponds to regular graphs, whereas $p=1/4$ corresponds to the most heterogeneous network that can be realized.  

Analyzing the giant component size after random attacks (Fig.~\ref{sxr}) shows that heterogeneity delays the breaking of the giant component ($r_{\rm b}(p) = (1 + 16p)/(2 + 16p)$), and leads to larger giant components after strong attacks. However, regular graphs perform better against weak attacks. Between these extremes, there is a transition region where different degrees of heterogeneity perform best.

%% file: sections/3_GT.tex

We now use a variational argument reminiscent of linear stability analysis in dynamical systems. We consider a generating function of the form \eq{\label{eqPerturb}G(x) = B(x) + \gamma D(x),}
where $B(x)$ is the generating function of the \lq base\rq~degree distribution, and $D(x) = \sum d_{k} x^{k}$ is the generating function of a perturbation to this base. We consider only perturbations that leave the norm and the mean degree invariant, which stipulates $D(1) = 0$ and $D'(1) = 0$.

We compute the response of the giant component to perturbations by first computing 
\eq{
\label{v_Ap}
v = Q_{\rm a}(v) = \frac{B'(A(v)) + \gamma D'(A(v))}{z}.
}
and hence 
\eq{
z \frac{{\rm d} v}{{\rm d} \gamma} = [B''(a) + \gamma D''(a)]A'(v) \frac{{\rm d}v}{{\rm d} \gamma} + D'(a),
}
where $a=A(v)$ for convenience. We then solve for 
\eq{
\label{v_gamma}
\frac{{\rm d} v}{{\rm d} \gamma} = \frac{D'(a)}{z - [B''(a) + \gamma D''(a)]c},}
where we used $A'(v)=c$. One can now compute 
\eqa{
\label{s_gamma}
\frac{{\rm d} s}{{\rm d} \gamma} &=& - \left[B'(a)+\gamma D'(a)\right]c\frac{{\rm d} v}{{\rm d} \gamma} - D(a) \\
&=& D'(a)\left(\frac{B'(a)+\gamma {D'}(a)}{B''(a) + \gamma D''(a)- z/c}\right)  - D(a).\label{lastbeforegam0}
}
Working in terms of $a$ simplifies these calculations considerably. We now use $a=cv+(1-c)$ in the form $c=(1-a)/(1-v)$ to eliminate the $c$. Moreover, to show that the $B$ corresponds to a locally optimal degree distribution, we consider the limit of $\gamma\to 0$ which yields
\eq{
\label{stabcond}
Y:=\left. \frac{{\rm d} s}{{\rm d} \gamma}\right|_{\gamma=0} = \frac{B'(a) D'(a)}{B''(a) - (z-B'(a))/(1-a)} - D(a). 
}
To demonstrate the local optimally of $B$, we need to show that $s$ can't be increased by any perturbation. For illustration, we consider again $z=3$. For small attacks ($r\to 0$), the regular graph ($B(x)=x^3$) is the optimal solution, as the giant component spans the entire network. Substituting $B$ in Eq.~(\ref{stabcond}) yields
\eq{
\label{attackonreg}
 Y(a) = \frac{a^2}{a-1}D'(a)  - D(a). 
}
In the limit of small attacks ($a\to 0$), we find $Y(a)=0$, consistent with the optimality of regular graphs against small attacks. Moreover, $Y'(0) = - D'(0) = - d_1$. This shows that, for small finite attacks, increasing (decreasing) $\gamma$ from 0 only increases the giant component size if the perturbation $D$ removes (adds) nodes of degree 1. However, this is unphysical as the number of nodes of degree 1 would become negative. Hence the regular graph remains optimal for sufficiently small attacks. 

To find the size of attack at which the optimality of the regular graph is lost, we need to consider specific perturbations. The simplest normalizable perturbations are the three-point perturbation, which only affects nodes of three degrees. 
Because $Y$ considers infinitesimal perturbations, we are operating within the realm of linear response, where a superposition of perturbations leads to a corresponding superposition of responses. 
Therefore, a degree distribution is locally optimal if it is optimal with respect to all 3-point perturbation.

Let's first consider the `1-3-4'-perturbation, $D(x)=x-3x^3+2x^4$, which removes nodes of degree 3 and creates nodes of degree 1 and 4. To find the critical value of $a$, where this perturbation starts increasing the giant component size, we substitute into Eq.~(\ref{attackonreg}) and then solve for $Y(a)=0$, which yields, $a \approx 0.542$. Using 
\eq{
\label{eqTranslation}
r = \frac{za - B'(a)}{z - B'(a)} = \frac{a}{1 + a},
}
this translates to $r\approx 0.351$.  
\begin{figure}[tb]
{\includegraphics[width=0.8\columnwidth]{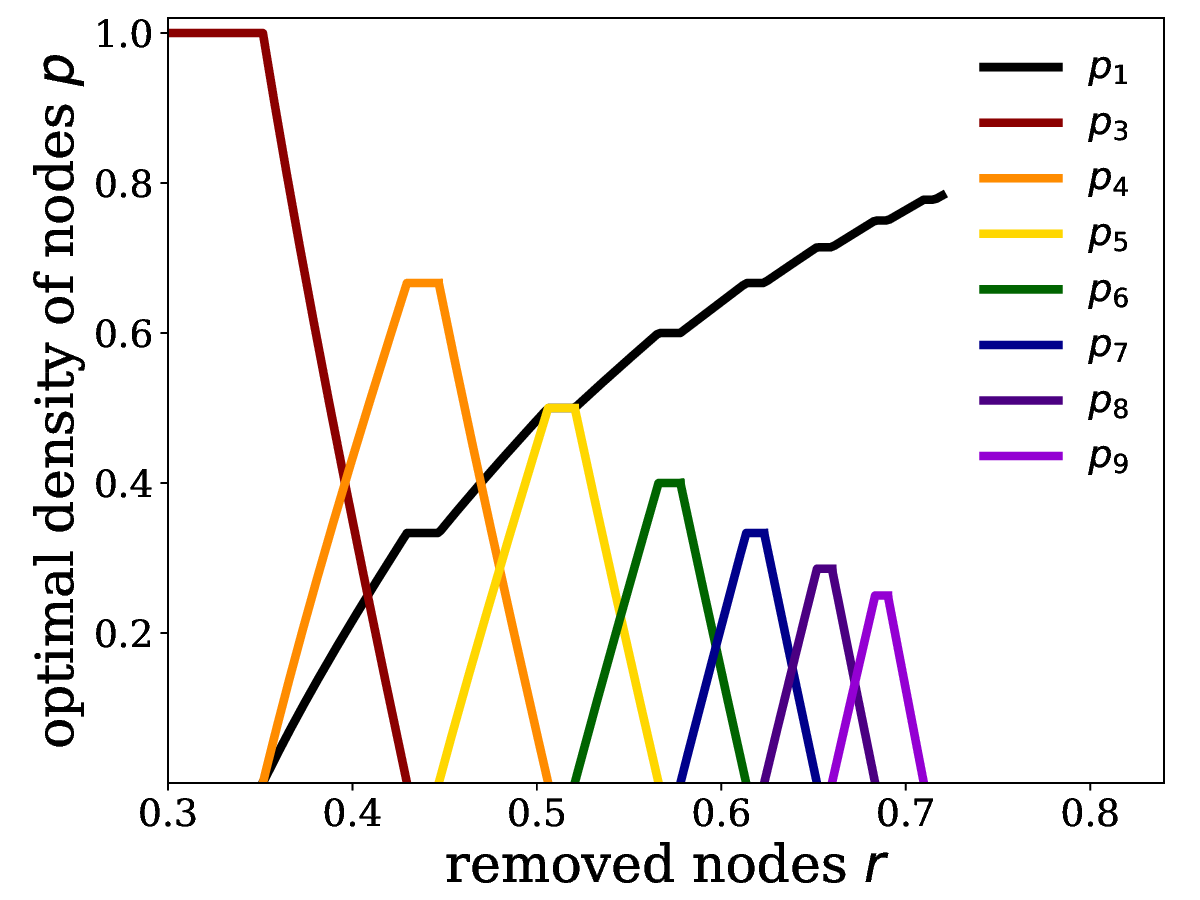}}
\caption{\label{ispt_s}Infinite sequence of continuous transitions. The elements of the degree distribution $p_k$ are order parameters of the system. As the size of the attack $r$ increases, an infinite sequence of phase transitions unfolds in which higher degrees successively appear.}
\end{figure}

We can now repeat this analysis with a general three point perturbation that removes nodes of medium degree $m$ to create nodes of a higher degree $h$ and lower degree $l$. This perturbation is generated by 
\eq{
D(x)=(h-m)x^l-(h-l)x^m+(m-l)x^h.
}
For example, considering a three-point perturbations with degrees $2,3,4$ yields a critical value of $a \approx 0.577$, and the perturbation with degrees $1,3,5$, has $a \approx 0.557$. We verified that among all three-point perturbations, 1-3-4 is the one that becomes beneficial at the smallest value of $a$. So, in a network with $z=3$, an initially regular degree distribution is optimal against attacks that remove less than 35\% of the nodes.
Subsequently, the optimal network is a mixture of the regular graph $B$ and the 1,3,4-perturbation $D$. In this range, we can find the optimal value of $\gamma$ by rewriting Eq.~(\ref{lastbeforegam0}) in the form
\eq{
\label{p_optimal}
\gamma(a) = \frac{D[z - B' - B''(1-a)] + D'B'(1 - a)}{D[D' + D''(1 -a)] - D'^{2}(1 - a)}.
}
Substituting $\gamma(a)$ back into Eq.~(\ref{eqPerturb}) gives us the optimal degree distribution in the region where nodes of degree 4 start to appear (Fig.\ref{ispt_s}). As the proportion of removed nodes increases, this solution becomes eventually unphysical when all nodes of degree 3 have been used up and the optimal number of nodes of degree 3 becomes negative. After this point, the optimal realizable solution is a network that consists entirely of nodes of degree 1 and 4. The corresponding degree distribution is generated by $B=(x+2x^4)/3$, we now use this distribution as our new base distribution and consider perturbations. Repeating the same steps as before, we arrive at  
\eq{
\frac{4a^2-2a+1}{8(a-1)}D'(a)-D(a)=0,
}
which is the equivalent to Eq.~(\ref{attackonreg}). We can now use the same approach as before to check the stability. Once we find the value of $a$, where a perturbation becomes advantageous, we identify the critical point in terms of $r$, using the appropriate relationship, which is $r=(8a^3-9a+1)/(8a^3-8)$ for the current $B$ (cf.~Eq.~\ref{eqTranslation}). This reveals that the network with node degrees 1 and 4 is the optimal solution between $r\approx 0.429$ and $r\approx 0.447$. 

We note that the elements of the degree distribution, $p_k$, are intensive and describe collective properties. They are thus well-defined order parameters, and the critical points where the parameters change non-smoothly correspond to continuous phase transitions.  

Repeating the same procedure as above reveals that network alternates between two types of phases as $r$ is increased. We think of the first type of phase as a `dynamic' phase, as the optimal degree distribution changes continuously with $r$ in these phases. The second type of phase is a static phase, where the degree distribution becomes fixed and does not change anymore until the next phase transition is encountered.

The network consisting only of nodes of degree 1 and 4 is such a static phase, where the proportion of nodes of degree 1 remains fixed at 1/3 regardless of $r$. However, beyond $r=44\%$, a phase transition occurs in which we enter a dynamic phase in which nodes of degree 4 are gradually replaced by degree 1 and 5. The dynamic phase ends at $r\approx 0.506$, where all nodes of degree 4 have been eliminated, and we enter a static phase with degrees 1 and 5. Eventually, ($r\approx 0.520$), we enter another dynamic phase were nodes of degree 6 start to appear (Fig.\ref{ispt_s}). 

As we increase $r$, the system transitions through a sequence of continuous phase transitions, which eventually ends in a static phase containing nodes of the highest realizable degree $k_{\rm max}$ as well as degree 1. We conjecture that in an infinite network this sequence of phase transitions is infinite and continues up the point $r=1$. This is supported by the observation of a scaling law of the form 
\eq{
c=1-r\sim k^{-\alpha}, 
}
that relates the surviving proportion of nodes after the attack, to highest degree that appears in the network.(Fig.~3). 

\begin{figure}[tb]
{\includegraphics[width=0.85\columnwidth]{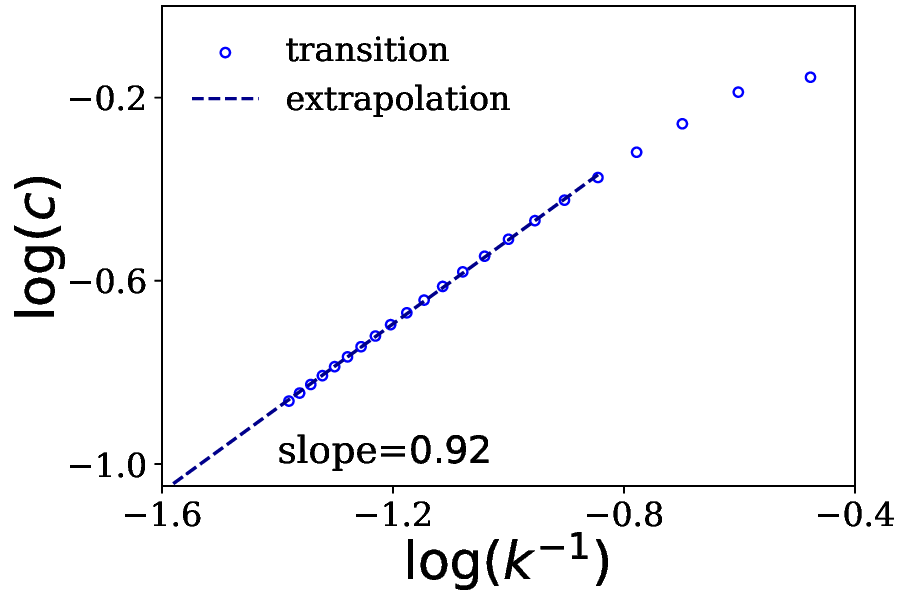}}
\caption{\label{figlog}Scaling of transition points. Shown are the transition points (circles) where degree $k$ first appears. The good agreement with a linear fit (line) suggest a power-law scaling behaviour of critical points.}
\end{figure}

So far, we have shown that optimizing the proportion of nodes in the giant component after the attack reveals a sequence of structural phase transitions. Let's now consider the related question of optimizing the proportion of links in the giant component. The probability that a node that we find at one end of the link is not otherwise connected to the giant component is $v$. The same is true for the nodes at both ends of the link, so the proportion of links in the giant component is $1-v^2$. Therefore, the degree distribution that maximizes the proportion of links in the giant component also minimizes $v$. From Eq.~(\ref{v_gamma}) we can see that the optimality condition is  $D'(a)=0$. This is interesting because it only depends on the perturbation $D$ and not the base $B$. Hence, the base distribution only determines whether a given perturbation is physically feasible. Once a perturbation is feasible and beneficial, it will immediately be applied with the maximal realizable $\gamma$. This precludes the existence of dynamic phases. Thus, each phase is now a static phase, and the transitions between these phases are discontinuous phase transitions.

The static phase with nodes of degree 1 and $k$ is generated by $B_{k}(a)= [(k -z)a + (z-1)a^{k}]/(k-1)$. We locate the phase transition points that connect these phases using the same procedure as before. This reveals that the transition between adjacent phases with nodes of degree $k$ and $k+1$, occurs when $B_{k}'(a)=B_{k+1}'(a)$. This correspondence between the points where the derivatives match and the phase transition point holds even in more constrained problems, where only certain degrees are allowed before the attack.

As a result of the analysis, we find a sequence of transitions that is reminiscent of the one observed before (Fig.~4). Again, the optimal network transitions through a sequence of static phases, where we find only nodes of degree 1 and $k$, where $k$ increases by one in every new static phase. However, in case of the maximization of links in the giant component the transitions between these static phases occur as discontinuous transitions.   
\begin{figure}[tb]
\includegraphics[width=0.8\columnwidth]{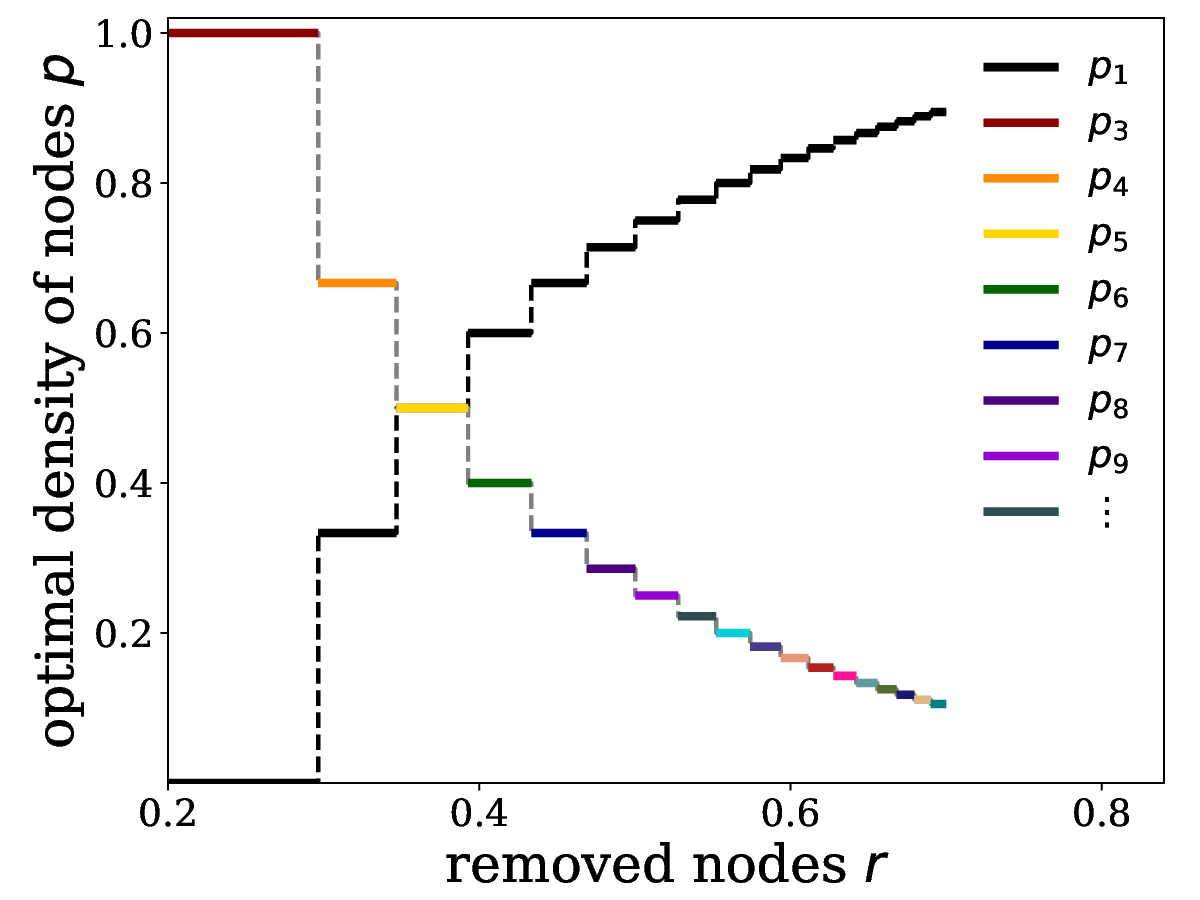}
\caption{\label{ispt_v}Infinite sequence of discontinuous transitions. Maximizing the proportion of links that are still in the giant component after an attack of size $r$ leads to a sequence of discontinuous phase transitions. In each phase, the degree distribution only consists of nodes of degree 1 and one other degree $k$ (before the attack). Nodes of degree $0$ are not allowed. }
\end{figure}

%% file: sections/4_Co.tex

In this Letter, we explored optimal network structures in configuration models that maximized the proportion of nodes or links in the giant component after a random attack of known size. The optimization was constrained by fixing the mean degree, demanding that each node has at least degree 1 before the attack. We showed that the optimal degree distributions depend on the size of the attack. As the proportion of removed nodes is increased, the optimal distribution passes through an infinite sequence of structural phase transitions. These transitions are continuous if the proportion of nodes in the giant component is maximized, but discontinuous if we maximize the proportion of links in the giant component instead. Given the attention that that such discontinuous transitions in networks have recently received, we were surprised to encounter such transitions in a foundational topic of network science such as network robustness. We believe that there is still much more to discover here, so further work in this area may be fruitful.